# Nuclear obscuration structure in Mrk 417 based on NuSTAR and Swift/BAT data
## Kompaniiets O.V., Vasylenko A.A.

Main Astronomical Observatory of the NAS of Ukraine, Kyiv, Ukraine

***kompaniets@mao.kiev.ua***

We present the results of X-ray spectral analysis of the nearby (z = 0.0327) Seyfert type 2 galaxy Mrk 417 observed by the NuSTAR (3-60 keV) and the Swift/BAT (14-150 keV). This analysis is based on the spectral data from a NuSTAR observation performed in February 2017 (ID 60061206002). In addition, the data from Swift/BAT 150-month survey were used. The NuSTAR spectrum shows a good fit with a baseline model composed of an absorbed cutoff power-law component and reflected emission from cold neutral medium. We found that spectrum is steep $\Gamma = 1.63^{+0.10}_{-0.11}$ and obscured by column density of $N_H = 3.22^{+0.41}_{-0.39} \cdot 10^{23}$ cm$^{-2}$. The narrow Fe K$_\alpha$ emission line at $6.33^{+0.12}_{-0.14}$ keV with an equivalent width (EW) of $115^{+2}_{-1}$ eV is also present.

We also applied a more physically motivated approach using the complex physical models MYTorus and BNTorus in combining NuSTAR data with Swift/BAT spectrum. MYTorus model used in "coupled" mode which implies an azimuthally symmetric obscured torus with a fixed opening angle $\theta_{tor} = 60°$. As a first step the relative normalizations $A_s$ and $A_L$ were fixed equal to 1. In this case we obtained a good fit with the power-law index $\Gamma = 1.6^{+0.09}_{-0.07}$, inclination angle $\theta_i = 87.5^{+peg}_{-7.7}$, and equatorial column density $N_{H (eq)} = 3.3^{+0.2}_{-0.2} \cdot 10^{23}$ cm$^{-2}$. After this, $A_s$ and $A_L$ were made free but still linked to each other. It means, that the region where formation scattered components of the spectrum and lines the same, but the obscuring geometry is different from the original. As a result, the photon index changed to $\Gamma_{MyTorus} = 1.68^{+0.09}_{-0.09}$, inclination angle is $\theta_i = 86.2^{+peg}_{-9.3}$. The equatorial column density was defined as $N_{H (eq)} = 3.40^{+0.04}_{-0.04} \cdot 10^{23}$ cm$^{-2}$ which correspond the line-of-sight column density of absorber $N_{H\,l.o.s} = 3.36^{+0.04}_{-0.07} \cdot 10^{23}$ cm$^{-2}$. The measured normalization parameters are $A_s = A_L = 0.68^{+0.8}_{-0.63}$. The last one could indicate either a deviation from the geometry adopted in the "coupled" mode or the presence of time delays between intrinsic and reflected components. If we assume that parameter $A_s$ is responsible only for the geometry, we can estimate the covering factor as $f_c = 0.5 \cdot A_s = 0.34^{+0.4}_{-0.32}$ which may be explained as a ring-like shape of torus.. We also applied BNTorus model as an additional check of this assumption. From the application of the BNTorus model we find that the photon index $\Gamma_{BNTorus} = 1.75^{+0.09}_{-0.09}$, absorption value $N_{H\,BNTorus} = 3.72^{+0.49}_{-0.39} \cdot 10^{23}$ cm$^{-2}$, inclination angle $\theta_i = 79.8^{+1.6}_{-13.5}$ and opening angle $\theta_{tor} = 73.1^{+10.9}_{-3.9}$. The covering factor was calculated as $f_c = \cos(\theta_{tor}) = 0.29^{+0.06}_{-0.19}$ which is consistent to the MYTorus result. The derived intrinsic X-ray luminosity in 2-10 keV is $\sim 3.16 \cdot 10^{43}$ erg/s. Considering the WISE data in Mid-Infrared (MIR) light, we also derived another estimation of $f_c \sim L_{12\,mkm}/L_{bol} \sim 0.12$.

The results obtained from the different models and multi-wavelength data allow us to suggest that obscuring material of Mrk 417 have preferably a ring-like geometry.

# Структура поглощающей среды в ядре галактики Mrk 417
## по данным NuSTAR и Swift/BAT

**Компаниец Е.В., Василенко А.А.**

*Представлены результаты анализа рентгеновского спектра Сейфертовской галактики 2-го типа Mrk 417 ($z \approx 0.0327$), по данным космических обсерваторий NuSTAR (3-60 кэВ) и Swift/BAT (14-150 кэВ). Спектр, полученный обсерваторией NuSTAR, хорошо описывается базовой степенной моделью, с нейтральным поглощением и дополнительной компонентой, обусловленной отражением от холодной нейтральной среды (степенной индекс $\Gamma = 1.63^{+0.10}_{-0.11}$, поглощение $N_H = 3.22^{+0.41}_{-0.39} \cdot 10^{23}$ см$^{-2}$). Наличие узкой эмиссионной линии Fe $K_\alpha$ с эквивалентной шириной $EW_{Fe\ K\alpha} = 115^{+2}_{-1}$ эВ свидетельствует об умеренной плотности среды, в которой формируется эта линия. Анализ данных NuSTAR вместе с данными Swift/BAT был выполнен с применением более сложных физических моделей MYTorus и BNTorus. В первом случае было получено значение степенного индекса $\Gamma_{MYTorus} = 1.68^{+0.09}_{-0.09}$ и величину поглощения на луче зрения $N_{H\ l.o.s} = 3.36^{+0.04}_{-0.07} \cdot 10^{23}$ см$^{-2}$. Применение модели BNTorus показывает значение степенного индекса $\Gamma_{BNTorus} = 1.75^{+0.09}_{-0.09}$ и поглощение на луче зрения $N_{H\ l.o.s.} = 3.72^{+0.49}_{-0.39} \cdot 10^{23}$ см$^{-2}$. Эти результаты позволяют оценить фактор перекрытия газопылевого тора $f_c \approx 0.29\text{-}0.34$ и определить величину исправленной на поглощение светимости $L_{intr\ 2-10\ кэВ} \sim 3.16 \cdot 10^{43}$ эрг/с. Дополнительный анализ данных наблюдений в ближнем ИК диапазоне указывает на то, что фактор перекрытия может составлять даже еще меньшую величину – $f_c \sim 0.12$. Полученные результаты свидетельствуют о том, что газопылевой тор, скорее всего, является сжатым по вертикали и его форма приближается к кольцеподобной структуре.*

***Ключевые слова: активные ядра галактик, газопылевой тор, Mrk 417, рентгеновское излучение***

1. *Введение.* Структура активных ядер галактик (далее АЯГ) описывается унифицированной схемой (далее УС АЯГ) [1]: в центре АЯГ находится сверхмассивная черная дыра с аккреционным диском, вокруг которых расположен газопылевой тор. Газопылевой тор является одним из ключевых элементов в УС и отвечает за классификацию сейфертовских галактик по отношению к углу наклона. Взаимодействие первичного излучения от короны аккреционного диска с веществом в торе приводит к образованию спектра отражения [2, 3, 4]. Характерными особенностями спектра отражения являются эмиссионные линии железа и других элементов, в частности нейтрального или ионизированного железа Fe K$_\alpha$ с энергиями 6.4 кэВ и 6.7 кэВ, а также характерный горб в непрерывном спектре на энергии ~ 20 - 40 кэВ. Суммарный спектр отраженной и первичной компонент, модифицированный поглощением, используется для определения характеристик и геометрии газопылевой среды АЯГ (например, [5, 6, 7, 8, 9, 10, 11]). Качество наблюдаемого спектра отражения в АЯГ значительно улучшилось после начала работы космической обсерватории NuSTAR [12], по причине появления возможности фокусирования излучения с энергией вплоть до 79 кэВ.

Учитывая характеристики газопылевой среды, АЯГ принято классифицировать по величине поглощения, поскольку эта среда играет значительную роль в формировании наблюдаемой формы спектра. Если величина столбцевой плотности $N_H < 1.25 \cdot 10^{24}$ см$^{-2}$ то АЯГ называют Комптон-тонким, а в случае когда $N_H \geq 1.25 \cdot 10^{24}$ см$^{-2}$ — Комптон — толстым[1].

Одним из кандидатов в Комптон-тонкие АЯГ является эллиптическая галактика класса E2 Mrk 417 [13] оптически классифицирована как сейфертовская галактика типа 2 [14] с красным смещением z = 0.0327 (~ 144 Мпк) [15]. В рентгеновском диапазоне наблюдалась космическими обсерваториями XMM-Newton, Suzaku, NuSTAR и Swift.

В работе [16] был проанализирован рентгеновский спектр Mrk 417, полученный обсерваторией XMM-Newton, для анализа которого применили ряд моделей. Первая модель является простым степенным законом с нейтральным поглощением. В результате такой аппроксимации получено необычно малое значение степенного индекса Γ=0.56 с плохой статистикой $\chi^2$/d.o.f. = 565/82. Вторая модель учитывала дополнительное поглощение с частичным перекрытием и присутствием эмиссионной линии в окрестности 6.5 кэВ, что привело к изменению степенного индекса Γ=2.25$^{+0.15}_{-0.17}$. Величина столбцевой плотности составила $N_H = 8.57^{+1.27}_{-0.69} \cdot 10^{23}$см$^{-2}$, эквивалентная ширина линии Fe K$_\alpha$ EW=114.77$^{+75.78}_{-80.97}$ эВ. В следующую модель авторы добавили еще один степенной закон с соответствующим дополнительным поглощением, предполагая, что поглощающая среда АЯГ имеет разную плотность. Основные параметры, полученные в этом приближении следующие: $N_{H1}$ =

---

1    http://mytorus.com/mytorus-manual-v0p0.pdf, см. раздел 2.1

$0.00^{+0.04}_{-0.00} \cdot 10^{22}$ см$^{-2}$, $N_{H2} = 5.41^{+2.50}_{-1.13} \cdot 10^{23}$ см$^{-2}$, степенные индексы[2] $\Gamma_1 = 2.36^{+0.16}_{-0.16}$, $\Gamma_2 = 0.88^{+0.98}_{-0.47}$, эквивалентная ширина линий железа EW=179.2$^{+87.71}_{-86.64}$ эВ. Использование модели с учетом отражения от нейтральной среды вместе с дополнительными данными Swift/BAT позволило оценить значение энергии экспоненциального завала $E_c = 106.7^{+112.1}_{-13.5}$ кэВ. При этом значение степенного индекса стало $\Gamma=0.73^{+0.34}_{-0.53}$, а столбцевая плотность - $N_H$ $1.82^{+0.9}_{-1.35} \cdot 10^{23}$см$^{-2}$, что оказалось меньше по сравнению с предыдущей моделью. Полученные значения степенного индекса не являются типичными для сейфертовских галактик, и это можно объяснить плохим качеством спектра, а также тем, что выбранная модель поглощения в данном случае является некорректной.

Наблюдательные данные космической обсерватории Suzaku были обработаны в работе [17]. Для расширения энергетического диапазона авторами также был использован спектр, полученный Swift/BAT во время 20-ти месячного обзора неба. В ходе проведенного анализа было установлено присутствие эмиссионной линии Fe K$_\alpha$ с наблюдательной энергией $E_{FeK_\alpha} = 6.37^{+0.25}_{-0.31}$ кэВ, эквивалентной шириной EW$_{FeK_\alpha} = 126^{+46}_{-46}$ эВ и поглощением в континууме $N_H = 4.57^{+0.01}_{-0.01} \cdot 10^{23}$ см$^{-2}$. В работе [18] представлен анализ данных обсерватории Suzaku вместе с данными 70-ти месячного обзора неба Swift/BAT в суммарном диапазоне 0.5 - 150 кэВ. Использованная авторами модель включала в себя два степенных закона с экспоненциальным завалом и поглощением, а также компоненту отражения от холодной нейтральной среды. В результате были получены значения степенного индекса $\Gamma=1.60^{+0.08}_{-0.09}$, поглощения $N_H = 4.45 \pm 0.32 \cdot 10^{23}$ см$^{-2}$, коэффициента отражения R=1.76$^{+0.24}_{-0.77}$ и эквивалентной ширины линии Fe K$_\alpha$ EW$_{FeK_\alpha} = 125^{+21}_{-21}$ эВ. Параметр отражения имеет достаточно большое значение, которое может быть частично объяснено фиксированным значением энергии экспоненциального завала на 300 кэВ и невысоким качеством наблюдательных данных. Спектральный анализ, проведенный на основе наблюдательных данных Suzaku и 9-ти месячного обзора АЯГ Swift/BAT в работе [19], с использованием модели учитывающей отражение от нейтральной среды, показывает величину степенного индекса $\Gamma=1.45^{+0.13}_{-0.11}$, поглощение $N_H = 4.45 \pm 0.32 \cdot 10^{23}$ см$^{-2}$ и верхний предел коэффициента отражения R = 0.19.

В данной работе представлены результаты спектрального анализа рентгеновского спектра Mkr 417, полученного космической обсерваторией NuSTAR вместе с данными 105-ти месячного обзора неба Swift/BAT. Целью работы было определение характеристик газопылевого тора Mrk 417, для чего спектр аппроксимировался разными моделями. Анализ кривых блеска на наличие переменности представлен в Разделе 3, спектральный анализ - в разделе 4. Дополнительная проверка полученных характеристик газопылевого тора на основе ИК-диапазона представлена в Разделе 5.

---

2    $\Gamma_1$ - описывает спектр до 3 кэВ, $\Gamma_2$ - после 3 кэВ, см. Рис. 4 (правая панель) в [16].

2. *Обработка данных.* Наблюдение галактики Mrk 417 космической обсерваторией NuSTAR (Nuclear Spectroscopic Array) было проведено 20.02.2017 (ObsID 60061206002). Подготовка данных к обработке проводилась с использованием программного обеспечения NuSTARDAS v.1.6.0 (NuSTAR Data Analysis Software package) пакета HEASOFT v.6.19. Очищенные файлы событий были получены с помощью стандартной подпрограммы nupipline с использованием базы калибровочных файлов CALDB v.20181022. Для получения спектра, кривой блеска источника и фона применялась подпрограмма nuproducts. Области источника и фона были выбраны для обоих детекторов FPMA и FPMB как круговые области радиусом 60" и 70" соответственно.

Для расширения спектрального диапазона был использован усредненный по времени спектр в диапазоне 14-195 кэВ, полученный космической обсерваторией Swift/BAT (Burst Alert Telescope) в результате 105-месячного обзора неба[3] [20]. В нашем случае диапазон энергий был ограничен до 150 кэВ, поскольку для спектрального бина с энергией 150-195 кэВ доминирует фоновое излучение.

3. *Кривые блеска.* Были построены и проанализированы кривые блеска Mrk 417 в диапазонах 3-10 кэВ и 10-60 кэВ, а также их отношение (Рис. 1). Каждая точка является сгруппированным бином данных с часовой шириной 1100 с. Визуально кривые блеска в обоих диапазонах демонстрируют небольшую переменность. Для математического анализа кривой блеска была использована программа FTOOLS *lcstats*. Аппроксимируя кривые блеска моделью, подразумевающую отсутствие вариаций (т. е. константой) в обоих диапазонах, были получены значения статистики $\chi^2$/d.o.f.=15.7/26 для 3-10 кэВ и $\chi^2$/d.o.f.=13/26 для 10-60 кэВ. Таким образом, данные наблюдения обсерваторией NuSTAR активного ядра Mrk 417 показывают отсутствие значимой переменности. Соответствующие средние значения скорости счета со стандартными отклонениями имеют значения 0.25±0.03 с$^{-1}$ для 3-10 кэВ и 0.24±0.03 с$^{-1}$ для 10-60 кэВ. Вследствие отсутствия значительных по амплитуде вариаций в кривых блеска, в дальнейшем использовался полный, усреднённый по времени, спектр от детекторов FPMA/FPMB.

---

4. *Спектральный анализ.* Спектры Mrk 417 анализировались с использованием программы XSPEC v.12.9.0u. Значения погрешностей соответствуют 90% доверительному интервалу для одного параметра ($\Delta\chi^2 = 2.71$). Для учета различий во взаимной калибровке детекторов FPMA, FPMB и Swift/BAT, а также вариации блеска в данных последнего, были введены константы $C_{FPMB}$ и $C_{BAT}$ (в моделях обозначено как constant). Полученные в процессе анализа значения $C_{FPMB}$ и $C_{BAT}$ близки к 1.

Детекторы FPMA и FPMB могут получать спектры до 79 кэВ, но в нашем случае после 60 кэВ спектр фона начинает доминировать над спектром источника. Таким образом, в спектральном анализе использовался диапазон 3-60 кэВ. Галактическое поглощение учтено во всех моделях применением модели `tbabs` [9] и равняется $N_{H\ Gal} = 1.88 \cdot 10^{20}$ см$^{-2}$ согласно данным Лейденского обзора Галактики [21][4]. Во время спектрального анализа мы придерживались последовательности, использованной в [5]. При вычислении светимостей использовались значения космологических параметров: $H_0 = 70$ км с$^{-1}$ Мпк$^{-1}$, $\Omega_\lambda = 0.73$ и $\Omega_M = 0.27$ [22].

4.1. *Феноменологические модели.* Сначала мы аппроксимировали наблюдательный спектр простыми феноменологическими моделями, первая из которых (модель А) включает в себя степенной континуум (`zpowerlaw`), поглощение нейтральной средой (`ztbabs`) и эмиссионную линию с гауссовским профилем (`zgauss`). Последнюю компоненту добавлено на основе предыдущих работ [17, 18, 16], согласно которым в спектре имеется эмиссионная линия Fe K$_\alpha$ на энергии $\sim 6.4$ кэВ.

Вид данной модели в формализме XSPEC:

Модель А = `constant*tbabs*ztbabs*(zpowerlaw+zgauss)`.

Модель демонстрирует хорошую подгонку ($\chi^2$/d.o.f. = 215/227) со значениями степенного индекса $\Gamma = 1.62^{+0.09}_{-0.09}$, столбцевую плотность $N_H = 3.21^{+0.49}_{-0.37} \cdot 10^{23}$ см$^{-2}$, энергией линии $E_{FeK\alpha} = 6.33^{+0.12}_{-0.14}$ кэВ, эквивалентной шириной $EW_{Fe\ K\alpha} = 107^{+3}_{-4}$ эВ (Рис. 2). Значение степенного индекса $\Gamma$ в пределах погрешности совпадает с результатом работы [18].

Эквивалентная ширина линии, в приближении ее формирования в "холодной" среде, связана пропорцией с $N_H$ и телесным углом $\Omega$ следующим образом [17]:

$$EW \sim 300 \frac{\Omega}{4\pi} \frac{N_H}{4 \cdot 10^{23} \text{см}^{-2}} \text{ эВ} \qquad (1)$$

Таким образом, можно использовать полученные значения столбцевой плотности $N_H$ и эквивалентной ширины для оценки значения $\Omega/4\pi$:

$$\frac{\Omega}{4\pi} \sim \left(\frac{EW}{300 \text{ эВ}}\right) \cdot \left(\frac{N_H}{4 \cdot 10^{23} \text{см}^{-2}}\right)^{-1} \qquad (2)$$

Поскольку рентгеновский спектр АЯГ может иметь экспоненциальное обрезание на энергии выше 100 кэВ, мы заменили простой степенной спектр (`zpowerlaw`) на степенной спектр с экспоненциальным обрезанием на высоких энергиях (`cutoffpl`) (модель B). Значение энергии экспоненциального завала для начала было зафиксировано на энергии 300 кэВ согласно [18, 16].

Вид модели B в формализме XSPEC:

Модель B = `constant*tbabs*ztbabs(cutoffpl+zgauss)`.

Лучшая подгонка в рамках Модели B показывает хорошую статистику $\chi^2/\text{d.o.f.}$ = 208/225. Параметры континуума — степенной индекс $\Gamma = 1.51^{+0.1}_{-0.1}$, столбцевая плотность $N_H$ = $3.00^{+0.40}_{-0.37} \cdot 10^{23}$ см$^{-2}$, а параметры линии $E_{\text{Fe K}\alpha} = 6.33^{+0.12}_{-0.14}$ кэВ, эквивалентная ширина $EW_{\text{Fe K}\alpha} = 115^{+2}_{-1}$ эВ. Полученные результаты в пределах погрешности совпадают с результатами, полученными при применении модели A. Энергия экспоненциального завала $E_{\text{cut-off}}$, при свободном варьировании не определяется. Значения параметров, полученные в результате применения обсуждаемых моделей, представлены в Таблице 1 (две первых строки).

4.2. *Учет компоненты отраженного излучения.* Наличие в спектре флуоресцентной эмиссионной линии Fe $K_\alpha$ свидетельствует о присутствии определенной доли отраженного излучения. Для того чтобы принять во внимание эту спектральную компоненту, мы выбрали модель `pexmon` [23]. Она описывает степенной спектр с экспоненциальным завалом вместе с отраженной компонентой от нейтральной среды в виде плоской поверхности и самосогласовано также включает в себя эмиссионные линии Fe $K_\alpha$, Fe $K_\beta$ и Ni $K_\alpha$. В модели `pexmon` свободным параметром является т. н. коэффициент относительного отражения R, который с геометрической точки зрения может быть определен как R = $\Omega$ / 2$\pi$, где $\Omega$ — это телесный угол отражателя, под которым облучается среда. Угол наклона отражающей среды был зафиксирован $\theta = 60°$, поскольку это значение выбирается типичным для галактик типа Сейферт 2 (например, [17, 6]).

Вид модели C в формализме XSPEC:

Модель C = `constant*tbabs*ztbabs(cutoff+pexmon)`.

Лучшая подгонка в рамках Модели C показывает хорошую статистику $\chi^2/\text{d.o.f.}$ = 212/226, степенной индекс $\Gamma = 1.63^{+0.10}_{-0.11}$, столбцевую плотность $N_H = 3.22^{+0.41}_{-0.39} \cdot 10^{23}$ см$^{-2}$ и коэффициент относительного отражения R = $0.23^{+0.21}_{-0.24}$. Ненулевое значение R свидетельствует о наличии небольшой компоненты отраженного излучения в наблюдаемом спектре. Полученные величины $\Gamma$ и $N_H$ в пределах погрешности не отличаются от результатов, полученных при применении моделей A, B. Значения спектральных параметров данной модели представлены в Таблице 1 (третья строка).

4.3. *Анализ спектра с использованием NuSTAR и SWIFT/BAT.* Получение оценки энергии экспоненциального завала требует расширения энергетического диапазона. С этой целью к спектральному анализу были добавлены данные Mrk 417 в диапазоне 14-150 кэВ, полученные космической обсерваторией Swift/BAT [20]. Применение к расширенному спектру модели C позволило получить значение $E_{cut-off} = 133^{+63}_{-34}$ кэВ. Значения степенного индекса $\Gamma = 1.57^{+0.18}_{-0.18}$ и коэффициента относительного отражения $R = 0.26^{+0.19}_{-0.22}$ при $\chi^2$/d.o.f. = 217/232 в пределах погрешностей не изменились. Поскольку модель поглощения `ztbabs` учитывает только фотоэлектрическое поглощение, но не учитывает комптоновское рассеяние, мы также применили более физическую модель `plcabs` [24]. Данная модель описывает распространение рентгеновского излучения от источника в сферически-симметричной среде с корректным учетом нерелятивистского комптоновского рассеяния и поглощения.

Результирующий вид модели в формализме XSPEC:

Модель D = `constant*tbabs(plcabs + pexmon)`.

Аппроксимация спектра моделью D (Рис. 3) показывает хорошую статистику $\chi^2$/d.o.f. = 220/232, значение степенного индекса $\Gamma = 1.57^{+0.18}_{-0.19}$, столбцовой плотности $N_H = 30.9^{+5.3}_{-4.7} \cdot 10^{22}$ см$^{-2}$ и энергии экспоненциального завала $E_{cut-off} = 142^{+230}_{-58}$ кэВ. Коэффициент относительного отражения немного увеличился и стал равен $R = 0.37^{+0.19}_{-0.23}$. Таким образом, величины спектральных параметров не изменились по сравнению с моделью C в пределах погрешностей. Однако отметим, что даже с применением улучшенной модели континуума `plcabs` и с учетом отражения от нейтральной среды, значение степенного индекса отличается от типичного значения АЯГ $\Gamma \sim 1.8$ являясь более пологим (см., например, [18, 6]). Такое различие может быть результатом некорректного учета отраженной компоненты спектра, особенно в случае, если считать, что области образования линии Fe $K_\alpha$ и комптоновского горба совпадают. Подход с использованием моделей D и F является консервативным подходом, при котором поверхность газопылевого тора описывается непрозрачной нейтральной плоскостью. Однако это сильно упрощенное приближение, поскольку согласно современным представлениям, газопылевая среда АЯГ имеет более сложную геометрию и не является абсолютно непрозрачной в рентгеновском диапазоне. Поэтому, было решено применить две модели, которые описывают более реалистические формы газопылевого тора.

4.3.1 *MYTorus.* MYTorus — это модель, которая была построена на основе Монте-Карло симуляций взаимодействия рентгеновского излучения с газопылевым тором в форме классического «бублика», значение фактора перекрытия которого фиксировано и составляет $f_c$

= 0.5 (угол раскрытия тора 60°) [25]. MyTorus состоит из нескольких компонент[5]: MYTorusZ, MYTorusS и MYTorusL. Первая компонента (MYTorusZ) описывает модификацию первичного излучения при прохождении через газопылевой тор. Вторая компонента (MYTorusS) представляет рассеянное и отражённое от стенок тора первичное излучение. Третья компонента (MYTorusL) характеризует излучение в линиях Fe $K_\alpha$, Fe $K_\beta$ и Ni $K_\alpha$, которые образуются в нейтральной среде вещества тора. Свободные параметры модели следующие: степенной индекс Γ, экваториальная столбцевая плотность ($N_{H(eq)}$) и угол наклона газопылевого тора $\theta_i$. Последний изменяется в пределах от 0° до 90°, где $\theta_i = 0°$ означает, что ориентация тора относительно наблюдателя "плашмя" (англ. "face-on») и $\theta_i = 90°$ - "с ребра" (англ. "edge-on "). В нашей работе модель MYTorus использовалась в т. н. "coupled" режиме (модель F). Другими словами, все параметры MYTorusS и MYTorusL приравнивались к MYTorusZ. Коэффициенты нормирования между соответствующими компонентами, $A_S$ и $A_L$, были зафиксированы и равнялись единице. Первичное излучение описывалось степенным законом. Экспоненциальный завал учитывался с помощью табличной модели MYTorusZ с фиксированным значением $E_{cut-off} = 160$ кэВ, которое близко к ранее полученному значению $E_{cut-off}$ с помощью моделей C и D.

Окончательный вид модели F = constant * tbabs (zpowerlaw * MYTorusZ + $A_s$ * MYTorusS + $A_L$* MYTorusL) или в формализме XSPEC:

Модель F = tbabs * constant (zpowerlw *
etable{ mytorus_Ezero_v00.fits} + constant *
atable{ mytorus_scatteredH160_v00.fits} + constant *
atable{mytl_V000010nEp000H}.

Применение модели F показало хорошую статистику $\chi^2$/d.o.f = 225/232, степенной индекс Γ = $1.65^{+0.09}_{-0.07}$ и угол наклона $\theta_i = 87.5^{+peg}_{-7.7}$ градусов. Величина экваториального поглощения $N_{H(eq)}$ = $3.3^{+0.2}_{-0.2} \cdot 10^{24}$ см$^{-2}$.

Изначально зафиксированный параметр $A_s$ включает в себя всю информацию о возможной переменности источника, отклонениях химического состава или геометрии газопылевой структуры от принятой в оригинальном варианте модели MYTorus. Отличие от единицы значения $A_s$ может свидетельствовать, в первую очередь, об отклонении геометрии от принятой в модели. Если интерпретировать $A_s$ исключительно как геометрический параметр, можно оценить фактор перекрытия как $f_c = 0.5 \cdot A_s$. Поэтому следующим шагом мы сделали параметры $A_S$ и $A_L$ свободными, но с условием $A_S = A_L$, подразумевая, что регион образования рассеянной компоненты спектра и линий один и тот же. Применение модели F в этом случае также показало хорошую статистику $\chi^2$/d.o.f = 225/231 (Рис. 4) , степенной индекс $\Gamma = 1.68^{+0.09}_{-0.09}$ и угол наклона $\theta_i = 86.2^{+peg}_{-9.3}$ градусов. Полученное значение параметра $A_s = 0.68^{+0.80}_{-0.63}$, что соответственно приводит к $f_c = 0.5 \cdot A_s = 0.34^{+0.40}_{-0.32}$. Этот результат можно интерпретировать как то, что газопылевой тор является более сжатым по вертикали и его форма приближается к кольцеподобной структуре. Величина экваториального поглощения по результатам аппроксимации $N_{H\,(eq)} = 3.40^{+0.04}_{-0.04} \cdot 10^{23}$ см$^{-2}$. Согласно формуле 1 в работе [25], чтобы получить значение поглощения на луче зрения, необходимо воспользоваться уравнением:

$$N_{H\,1.o.s} = N_{H(eq)}\,(1 - 4\cos^2\theta_i)^{1/2}, \qquad (3)$$

что для нашего случая дает $N_{H\,1.o.s} = 3.36^{+0.04}_{-0.07} \cdot 10^{23}$ см$^{-2}$. Вычисленное значение сравнимо с такими же полученными в предыдущих моделях. Величина $N_{H\,1.o.s}$ может быть переведена в оптическую толщину, используя:

$$\tau_s \sim x\sigma_T N_H \sim 0.809 N_{24} \sim 0.27, \qquad (4)$$

где, $\sigma_T$ - томсоновское сечение рассеяния, x — среднее количество электронов на один атом водорода, $N_{24}$ - столбцевая плотность, выраженная в единицах $10^{24}$ см$^{-2}$. Таким образом, ядро галактики Mrk 417, с точки зрения значения $N_H$, является Комптон-тонким (Compton-thin). Угол между осью тора и наблюдателем $\theta \approx 86°$ свидетельствует о том, что тор наблюдается "с ребра". Исправленная на поглощение светимость ядра Mrk 417 в диапазоне энергий 2 - 10 кэВ имеет значение $L^{intr}_{2-10кэВ} = 3.16 \cdot 10^{43}$ эрг/с,    что примерно в 1.4 раза больше, чем значения, полученные при использовании предыдущих моделей A, B, C, D (см. соответствующие столбцы в Таб. 1, 2). Такое различие объясняется разной геометрией поглощающей среды, принятой при спектральном анализе.

4.3.2 *BNTorus*. BNTorus — это модель, созданная на основе Монте-Карло симуляций, которая описывает взаимодействие рентгеновского излучения от центрального точечного источника с газопылевой средой [26]. Эта среда имеет вид сферы, модифицированной двумя полярными коническими пустотами. В модели учтено комптоновское рассеяние, фотоэлектрическое поглощение и флуоресцентные эмиссионные линии железа. Свободными параметрами модели является поглощение на луче зрения $N_H$, степенной индекс $\Gamma$, угол раскрытия тора $\theta_{tor}$ (варьируется в пределах от 25.8º до 84.3º) и угол, под которым ориентирована экваториальная плоскость тора по отношению к наблюдателю $\theta_{incl}$ (варьируется в пределах от 18.2º до 87.1º). Энергия экспоненциального завала не варьируется и зафиксирована $E_{cut-off} = 300$ кэВ.

Вид модели в формализме XSPEC:

Модель K = constant*tbabs*atable{torus1006.fits}.

Применение модели демонстрирует хорошую статистику $\chi^2$/d.o.f. = 224/232, степенной индекс $\Gamma = 1.75^{+0.09}_{-0.09}$, поглощение $N_H = 3.72^{+0.49}_{-0.39} \cdot 10^{23}$ см$^{-2}$. Полученное значение угла наклона экваториальной плоскости тора $\theta_{incl} = 79.8^{+1.6}_{-13.5}$ градусов в пределах погрешности совпадает со значением, полученным при применении модели MYTorus. Угол раскрытия тора $\theta_{tor} = 73.1^{+10.9}_{-3.9}$ градусов, как и в предыдущей модели, можно связать с фактором перекрытия этого тора $f_c = \cos\theta_{tor} = 0.29^{+0.06}_{-0.19}$, что сравнимо с $f_c$ в модели MYTorus. Таким образом, модель BNTorus также указывает на возможно близкую к кольцеподобной структуру газопылевой среды. Полные результаты применения модели представлены в Таблице 2.

Поскольку энергия экспоненциального завала, которая была определена в предыдущих моделях как $E_{cut-off} \sim 140$ кэВ и значительно отличается от принятого значения $E_{cut-off}$ в BNTorus, была дополнительно выполнена аппроксимация спектра с измененной энергией экспоненциального завала, для чего модель K умножалась на компоненту `zhighect`. При этом было рассмотрено два варианта:

1. Значение $E_{cut-off} = 140$ кэВ зафиксировано. В таком случае сразу получаем уменьшение степенного индекса и концентрации: $\Gamma = 1.56^{+0.07}_{-0.13}$, $N_H = 3.25^{+0.61}_{-0.85} \cdot 10^{23}$ см$^{-2}$. Углы $\theta_{tor}$ и $\theta_{incl}$ не претерпевают изменений. Значение статистики $\chi^2$/d.o.f. = 218/231.

2. $E_{cut-off}$ варьируется вместе с другими параметрами. Результирующая подгонка показывает статистику $\chi^2$/d.o.f. =217/230 и значения параметров $\Gamma = 1.44^{+0.13}_{-0.19}$, $N_H = 2.77^{+0.42}_{-0.19} \cdot 10^{23}$ см$^2$, $E_{cut-off} = 90^{+85}_{-25}$ кэВ.

Изменение энергии экспоненциального завала до 90 кэВ и степенного индекса до ~1.4 свидетельствует о вырождении параметров и некорректности модели. Причиной такого вырождения может являтся то, что в BNTorus отсутствует разделение на отдельно варьированные компоненты спектра по примеру модели MYTorus.

5. *Ближнее инфракрасное излучение и фактор перекрытия.* В ближнем инфракрасном (англ. - mid-Infrared radiation - MIR) диапазоне основными источниками излучения в активных ядрах галактик являются либо околоядерные зоны активного звездообразования, либо газопылевой тор. В последнем случае высокоэнергетическое излучение от центра АЯГ поглощается и переизлучается в инфракрасном диапазоне и тогда MIR может быть хорошим показателем собственной светимости АЯГ в рентгеновском диапазоне [27, 28].

Но сначала нужно проверить, действительно ли излучение MIR для Mrk 417 в данном диапазоне вызвано излучением от центра АЯГ. Для этого из каталога космической обсерватории WISE (Wise-Field Infrared Survey Explorer) [29] были взяты значения магнитуд в фильтрах $W_1$ (3.4 мкм) и $W_2$ (4.6 мкм). Согласно критерию из работы [30], если $W_1$ - $W_2$> 0.8, то доминирующей причиной ближнего инфракрасного излучения является активность ядра галактики. В нашем случае, $W_1$ - $W_2 = 11.042 - 10.105 \approx 0.94$. Таким образом, данные инфракрасного диапазона для Mrk 417 действительно можно использовать для получения независимой оценки светимости АЯГ в рентгеновском диапазоне. Воспользовавшись корреляционной зависимостью между светимостью $L_{2-10\text{кэВ}}^{\text{intr}}$ и $L_{12\text{ мкм}}$ [27]:

$$\log\frac{L_{12\text{ мкм}}}{10^{43}\text{эрг/с}} = (0.33 - 0.04) + (0.97 - 0.03)\cdot\log\frac{L_{2-10\text{ кэВ}}^{\text{intr}}}{10^{43}\text{эрг/с}}, \qquad (5)$$

мы получим, каким должно быть ожидаемое значение светимости в диапазоне 2-10 кэВ. Согласно данным обсерватории WISE для Mrk 417, значение $\log L_{12\text{ мкм}} = 43.58 \pm 0.01$ [16]. Тогда использование зависимости $L_{2-10\text{кэВ}}^{\text{intr}} - L_{12\text{ мкм}}$ показывает ожидаемое $L_{2-10\text{кэВ}}^{\text{intr}} \sim 1.83\cdot10^{43}$ эрг/с. Полученная же светимость при аппроксимации спектра моделью MYTorus $L_{2-10\text{кэВ}}^{\text{intr}} = 3.16\cdot10^{43}$ эрг/с, а это означает, что фактическая светимость, исправленная на внутреннее поглощение, примерно в 1.7 раза превышает ожидаемое значение. Принимая во внимание что, светимость в рентгеновском диапазоне является модельно зависимой величиной, полученная разница не является большой. Мы дополнительно обсудим этот результат в разделе 6.

Как было показано в предыдущей главе, меньшее за единицу отношение нормировок внутреннего континуума к рассеянному/отраженному $A_\text{S}$, при исключительно геометрической интерпретации, может быть использовано как указатель на, возможно, кольцеподобную сплющенную структуру газопылевой среды. Данный вывод также можно косвенно проверить, используя соотношение между светимостью тора (в виде светимости на 12 мкм) и болометрической светимостью АЯГ (согласно работам [31, 32]. Болометрическую светимость можно получить, умножив $L_{2\text{-}10\text{кэВ}}$ на 10 [33], тогда $L_\text{bol} \approx 3.2\cdot10^{44}$ эрг/с. Для Mrk 417, таким образом, $f_\text{c} = L_\text{тора}/L_\text{bol} = L_{12\text{ мкм}}/L_\text{bol} \sim 0.12$. Это значение согласуется с нашим предположением о кольцеподобном торе несмотря на то, что полученная величина приблизительно в 3 раза меньше такой же из рентгеновских данных[6]. В случае малого $f_\text{c}$ значительная часть рентгеновского излучения будет покидать центр АЯГ, не пересекая область газопылевого тора. В результате вклад отраженной компоненты в наблюдаемый спектр будет небольшим, что, соответственно, отражается на слабости линии железа Fe $K_\alpha$, а также на небольшом значении коэффициента относительного отражения R в использованной модели pexmon. На основе данных инфракрасного диапазона можно оценить величину поглощения, используя корреляционную зависимость [28]:

$$\log\left(\frac{N_H}{\text{см}^2}\right) = (14.37 \pm 0.11) + (0.67 \pm 0.11)\cdot\log\left(\frac{F_{12\text{ мкм}}}{F_{2-10\text{ кэВ}}^{\text{obs}}} \cdot \frac{\text{эрг с}^{-1}\text{см}^{-2}}{\text{мЯн}}\right). \qquad (6)$$

Согласно данным обсерватории IRAS, поток $F_{12\text{ мкм}} = 0.132$ Ян [34]. Наблюдательный поток в рентгене $F_{2-10\text{кэВ}}^{\text{obs}} = 3.16\cdot10^{-12}$ эрг с$^{-1}$/см$^2$. В этом случае использование зависимости «$N_\text{H}$

---

6 Использование отношения $L_{12\text{ мкм}}/L_\text{bol}$ для определения $f_\text{c}$ является очень грубым приближением, поскольку необходимо знать точное распределение плотности в газопылевом торе, его возможную пространственную анизотропию, а также полностью исключить влияние внеядерных источников ближнего ИК диапазона (см. например раздел 4.3 в [31], раздел 3.3 в [30]).

$- F_{12\text{ мкм}}/F_{2-10\text{ кэВ}}^{\text{obs}}$» показывает ожидаемое значение $N_H \sim 3.16 \cdot 10^{23}$ см$^{-2}$, что совпадает с результатами, полученными при применении феноменологических моделей и MYTorus.

6. *Результаты и обсуждения*. Мы представили результаты спектрального анализа данных галактики Mrk 417 в диапазоне 3-60 кэВ, полученных космической обсерваторией NuSTAR вместе с данными 14-150 кэВ из 105-ти месячного обзора неба Swift/BAT. Эта галактика была классифицирована как кандидат в Комптон-тонкие АЯГ по результатам работы [18]. Использование степенного закона с поглощением в спектральном анализе показало степенной индекс $\Gamma \sim 1.6$ и $N_H \sim 3 \cdot 10^{23}$ см$^{-2}$. Добавление модели отражения в пределах погрешностей не изменяет параметры континуума, но при этом показывает небольшое наличие в спектре отраженной компоненты $R = 0.37^{+0.19}_{-0.23}$ и экспоненциального завала с энергией $E_{\text{cut-off}} = 142^{+230}_{-58}$ кэВ. Полученные значения спектральных параметров отличаются от результатов, которые были представлены ранее в работах [17, 18, 16, 19].

С целью получения дополнительных характеристик газопылевой среды в Mrk 417 также было рассмотрено физические модели MYTorus и BNTorus. Степенной индекс по результатам применения этих моделей, $\Gamma_{\text{MYTorus}} = 1.68^{+0.09}_{-0.09}$ и $\Gamma_{\text{BNTorus}} = 1.75^{+0.09}_{-0.09}$, имеет значение типичное для АЯГ. Также с помощью модели MYTorus было установлено, что газопылевая среда может быть описана однородным тором с экваториальным поглощением $N_{H\,\text{MYTorus}} = 3.4^{+0.4}_{-0.4} \cdot 10^{23}$ см$^{-2}$ наблюдаемым под углом $\theta_i = 86.2^{+peg}_{-9.3}$ градусов. Возможность интерпретировать коэффициент нормировки $A_s$ как геометрический параметр позволило оценить фактор перекрытия тора $f_c = 0.34^{+0.40}_{-0.32}$. Модель BNTorus использовалась для дополнительной оценки угла наблюдения и фактора перекрытия. По результатам применения этой модели, угол наклона экваториальной плоскости тора $\theta_{\text{incl}} = 79.8^{+1.6}_{-13.5}$ в пределах погрешности совпадает со значением, полученным при использовании модели MYTorus. Угол раскрытия тора $\theta_{\text{tor}} = 73.1^{+10.9}_{-3.9}$ градусов, что эквивалентно фактору перекрытия $f_c = 0.29^{+0.06}_{-0.19}$, сравнимому с $f_c$ модели MYTorus. Основываясь на полученном значении $f_c$, можно предположить, что тор является сжатым по вертикали и его форма приближается к кольцеподобной структуре. Поскольку газопылевая среда также переизлучает в ближнем инфракрасном диапазоне, были использованы данные наблюдений космической обсерватории WISE для дополнительной проверки результатов, полученных в рентгеновском анализе. С этой целью использовалась корреляционная зависимость $L_{2-10\text{ мкм}}^{\text{intr}} - L_{12\text{ мкм}}$, которая позволяет определить ожидаемое значение светимости в рентгене $L_{2-10\text{кэВ,MIR}}^{\text{intr}} \sim 1.83 \cdot 10^{43}$ эрг/с, что примерно в 1.7 меньше светимости, полученной при аппроксимации спектра моделью MYTorus $- L_{2-10\text{кэВ}}^{\text{intr}} = 3.16 \cdot 10^{43}$ эрг/с. Здесь важно отметить, что анализ излучения ближнего инфракрасного диапазона позволяет делать выводы только о пылевых структурах, в то время как на рентгеновские данные также влияет газ, локализированный внутри радиуса сублимации пылевых частиц. Более того, пространственное отношение количества газа и пыли может быть неоднородным вдоль луча зрения. В отношении последнего, подчеркнем тот факт, что в трех работах [17, 18, 19] было получено систематически большие значения столбцевой плотности (в среднем на $\sim 55\%$), при этом были использованы данные, включающие диапазон 0.5-3.0 кэВ, более чувствительный к особенностям поглощающей среды[7].

Как дополнительная оценка фактора перекрытия использовалось соотношение между светимостью на 12 мкм и болометрической светимостью: $f_c \approx L_{12\ мкм}/L_{bol} = 0.12$, что согласовывается с предположением о кольцеподобной структуре тора. Также это может объяснить полученное малое значение параметра отражения $R \approx 0.37$, так как в предположении кольцеподобной структуры газопылевой среды большая часть рентгеновского излучения будет покидать центр без взаимодействия с этой средой.

Главная астрономическая обсерватория НАН Украины, ул. Академика Заболотного, 27, г. Киев, Украина, e-mail: kompaniets@mao.kiev.ua.

## ЛИТЕРАТУРА

---

7   Имеется в виду, что в мягком рентгеновском диапазоне проявляются сигнатуры поглощения газовой средой (см., например, [36])

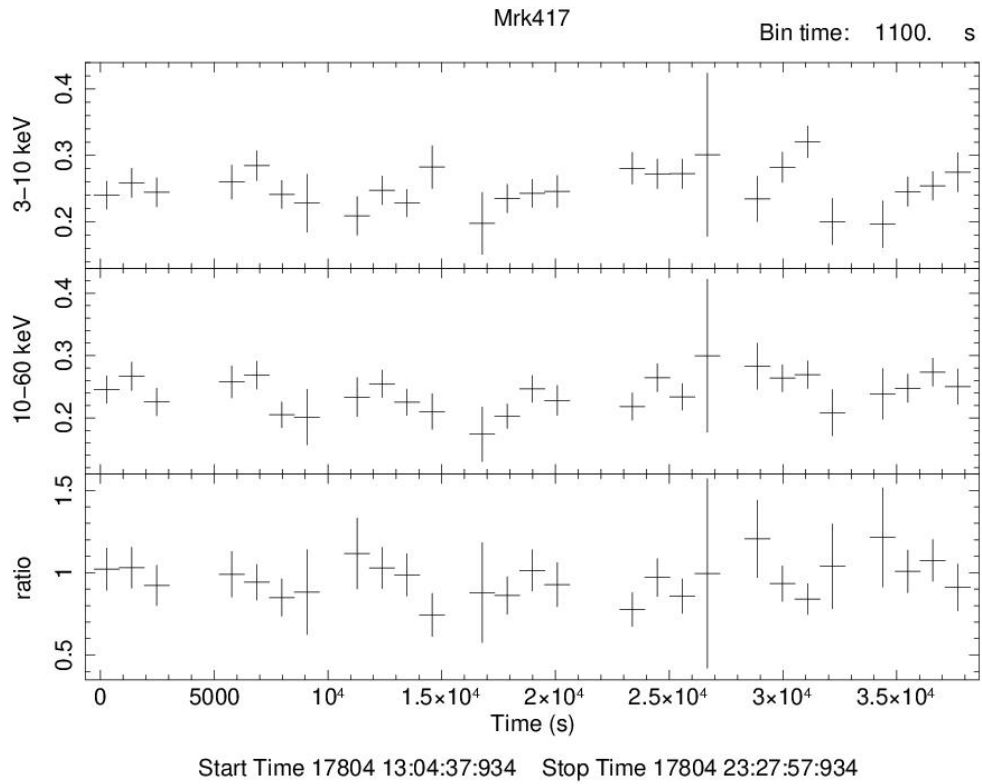

Fig.1 FPMA + FPMB luminosity curves in the energy ranges of 3 - 10 keV (upper panel), 10 - 60 keV (middle panel), and their ratio (lower panel).

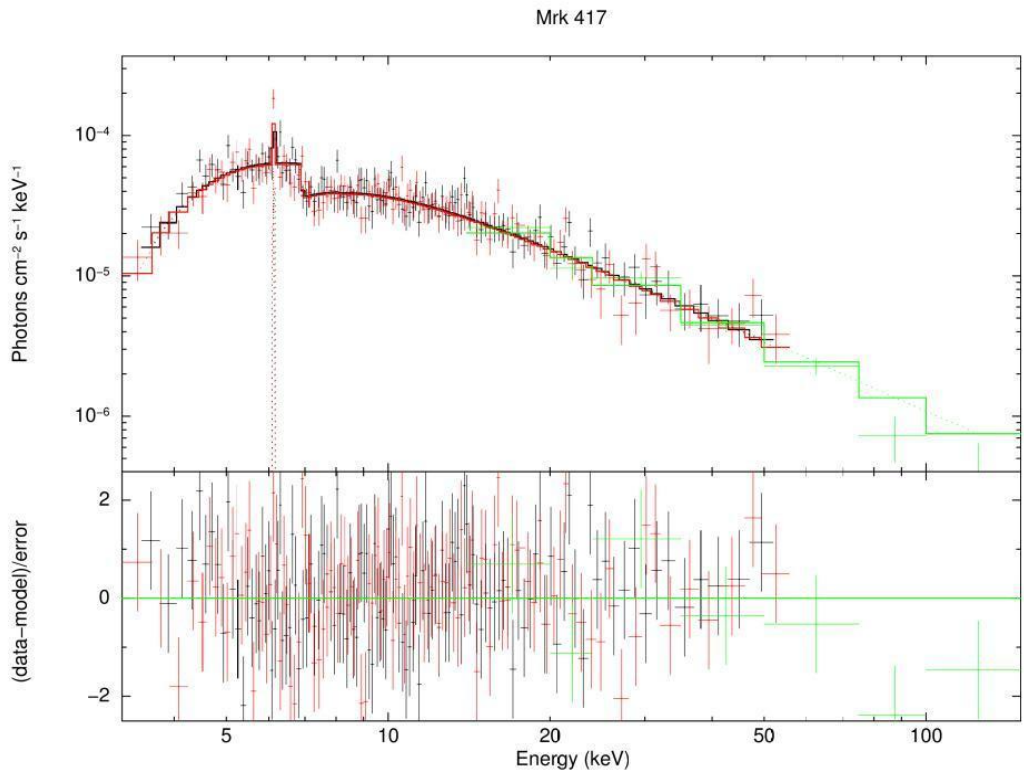

Fig.2 The best-fit folded spectrum with model A. In the lower panel the residuals are shown. Black and red colors refer to NuSTAR data, green to Swift/BAT.

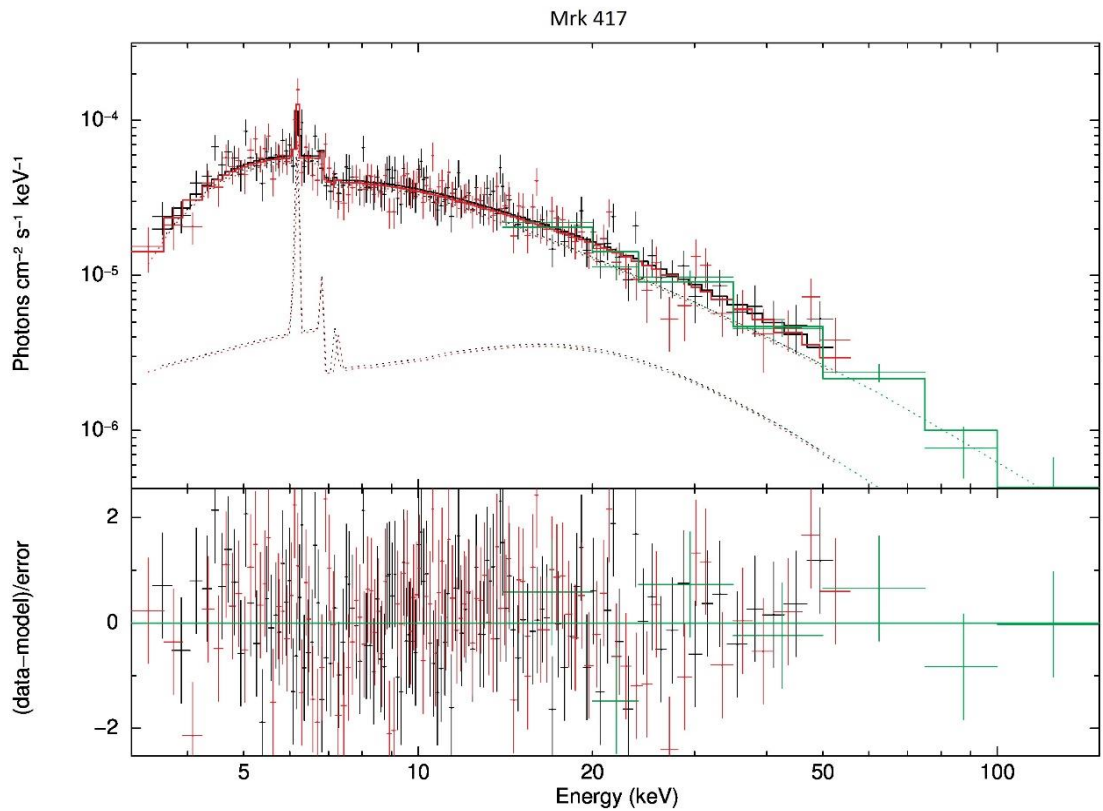

Fig.3 The best-fit folded spectrum with model D. In the lower panel the residuals are shown. Black and red colors refer to NuSTAR data, green to Swift/BAT.

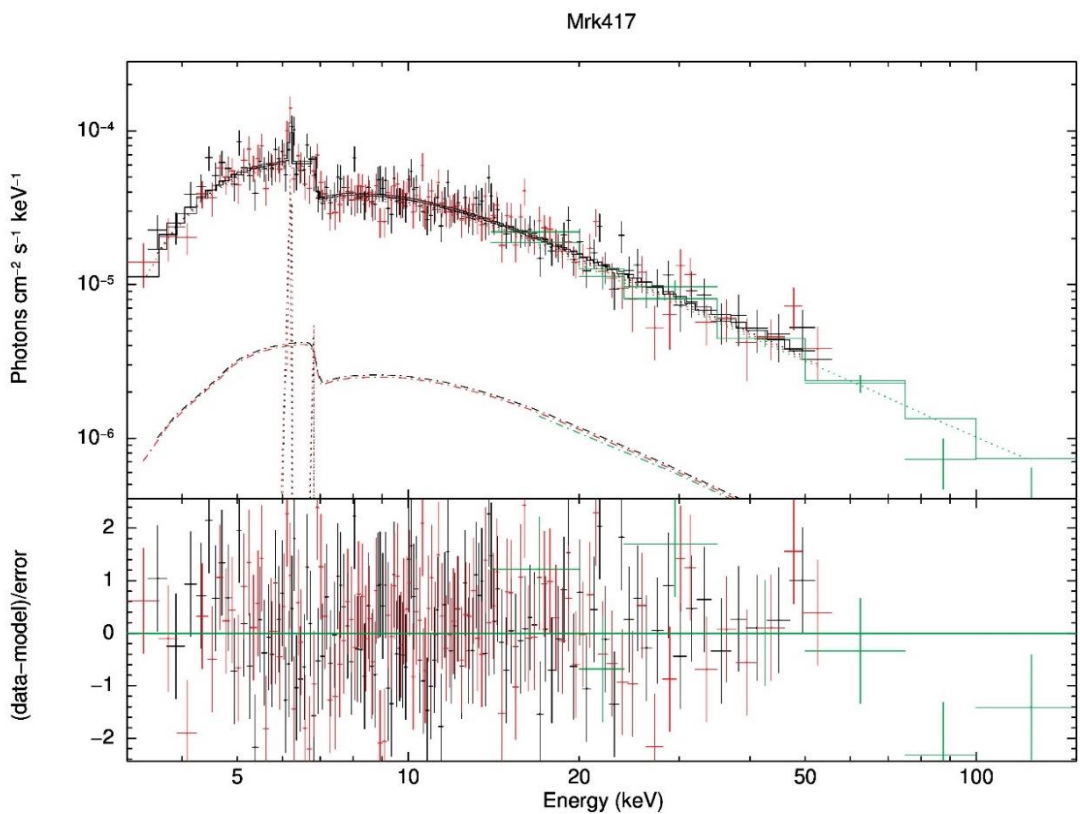

Fig.4 The best-fit folded spectrum with MYTorus model . In the lower panel the residuals are shown. Black and red colors refer to NuSTAR data, green to Swift/BAT.

Table 1. Spectral parameter values for best fit of the Mrk 417 spectrum according to NusTAR data.

| Baseline model | $\Gamma$ | $N_{\mathrm{H}}$, $10^{23}\,\mathrm{cm}^{-2}$ | R | E, keV | $\Theta\mathrm{i}$, deg | $L^{\mathrm{intr}}_{2-10\mathrm{keV}}$ erg/s | $L^{\mathrm{intr}}_{10-40\mathrm{keV}}$ erg/s | $\chi^2$/d.o.f |
|---|---|---|---|---|---|---|---|---|
| zpowerlaw | $1.62^{+0.09}_{-0.09}$ | $3.21^{+0.49}_{-0.37}$ | - | $6.33^{+0.12}_{-0.14}$ | - | $2.45\cdot10^{43}$ | $3.70\cdot10^{43}$ | 215/227 |
| cutoffpl | $1.51^{+0.1}_{-0.1}$ | $3.0^{+0.40}_{-0.37}$ | - | $6.33^{+0.12}_{-0.14}$ | - | $2.23\cdot10^{43}$ | $3.58\cdot10^{43}$ | 208/225 |
| cutoffpl+ pexmon | $1.63^{+0.10}_{-0.11}$ | $3.22^{+0.41}_{-0.39}$ | $0.23^{+0.21}_{-0.24}$ | - | $60^{(f)}$ | $2.34\cdot10^{43}$ | $3.19\cdot10^{43}$ | 212/226 |

\* f — fixed values

Table 2. Spectral parameter values for best fit of the Mrk 417 spectrum according to NuSTAR and Swift/BAT data.

| Baseline model | $\Gamma$ | $N_{\mathrm{H}}$, $10^{23}\,\mathrm{cm}^{-2}$ | R | $\Theta_{\mathrm{tor}}$, deg | $\Theta_{\mathrm{i}}$, deg | $E_{\mathrm{cut}}$, keV | $L^{\mathrm{intr}}_{2-10\mathrm{keV}}$ erg/s | $L^{\mathrm{intr}}_{10-40\mathrm{keV}}$ ergr/s | $\chi^2$/d.o.f |
|---|---|---|---|---|---|---|---|---|---|
| cutoffpl+ pexmon | $1.57^{+0.18}_{-0.18}$ | $3.12^{+0.49}_{-0.46}$ | $0.26^{+0.19}_{-0.22}$ | - | $60^{(f)}$ | $133^{+63}_{-34}$ | $2.22\cdot10^{43}$ | $3.25\cdot10^{43}$ | 217/238 |
| plcabs+ pexmon | $1.57^{+0.18}_{-0.19}$ | $3.09^{+0.53}_{-0.47}$ | $0.37^{+0.19}_{-0.23}$ | - | $60^{(f)}$ | $142^{+230}_{-58}$ | $2.17\cdot10^{43}$ | $3.13\cdot10^{43}$ | 220/232 |
| MyTorus | $1.68^{+0.09}_{-0.09}$ | $3.4^{+0.4}_{-0.4}$ | - | - | $86.2^{+p}_{-9.3}$ | $160^{(f)}$ | $3.16\cdot10^{43}$ | $4.37\cdot10^{43}$ | 225/232 |
| BNTorus | $1.75^{+0.09}_{-0.09}$ | $3.72^{+0.49}_{-0.39}$ | - | $73.1^{+10.9}_{-3.9}$ | $79.8^{+1.6}_{-13.5}$ | $300^{(f)}$ | - | - | 224/232 |

\* f — fixed values